# A Digital Twin for Individualized Treatment Effects of Non-Invasive Respiratory Support Strategies (DINIRS)

Md Fantacher Islam, MS[1], Jarrod Mosier, MD[2], Vignesh Subbian, PhD[1]
[1]College of Engineering, The University of Arizona, Tucson, AZ, USA
[2]College of Medicine - Tucson, The University of Arizona, Tucson, AZ, USA

**Abstract**

***Objective:*** Choosing between noninvasive respiratory support (NIRS) and invasive mechanical ventilation (IMV) for acute respiratory failure is a time-sensitive decision. Although guidelines provide population-level guidance, it remains unclear who benefits more from NIRS than IMV. The standard outcome, ventilator-free days at 28 days (VFD-28), scores death and prolonged ventilation equally, and current estimators do not distinguish between them. We developed and validated a censoring-aware Digital Twin framework for Individualized Treatment Effects of Non-Invasive Respiratory Support (DINIRS) to estimate individualized treatment effects (ITEs) that capture both mortality and ventilation duration.

***Materials and Methods:*** We emulated a target trial in 5,336 MIMIC-IV patients and trained DINIRS on 23 baseline clinical variables measured during the first 24 ICU hours. A transformer encoder with a survival attention gate decomposed VFD-28 into survival probability and conditional ventilation duration. A cross-fitted, doubly robust learner estimated ITEs. We externally validated DINIRS in 2,540 patients from the multi-site eICU-CRD dataset without retraining.

***Results:*** The DINIRS policy achieved a mean benefit of 2.07 ventilator-free days per patient (207 per 100 patients) compared with observed practice. Predicted NIRS benefit was higher among patients with less organ dysfunction (88.4% versus 49.0%) and persisted across hypoxemia severity. External validation reproduced this pattern.

***Discussion*****:** The NIRS benefit stemmed from shorter ventilation among survivors rather than from reduced mortality, indicating that avoiding intubation-associated complications was the primary mechanism.

***Conclusion*****:** Prospective validation is needed before these estimates inform treatment decisions. The decomposition framework can be extended beyond respiratory support to any zero-inflated composite outcome.

## 1.0 Background and Significance

Choosing the appropriate respiratory support for patients with acute respiratory failure in the emergency department (ED) or intensive care unit (ICU) is a complex, time-sensitive clinical decision that can affect morbidity and mortality.[1,2] Clinicians must decide whether to initiate non-invasive respiratory support (NIRS), which includes high-flow nasal oxygen (HFNO) and noninvasive positive pressure ventilation (NIPPV) modalities (e.g., continuous positive airway pressure [CPAP] and bilevel positive airway pressure [BPAP]), or to proceed directly to endotracheal intubation and invasive mechanical ventilation (IMV).[3,4] This decision is often complicated by diverse underlying pathophysiologies, such as pulmonary edema, obstructive airway disease, and acute lung injury, as well as patient-related factors such as body habitus, comorbidities, and mental status.[5,6] Because these heterogeneous syndromes respond variably to different support strategies, clinicians face substantial uncertainty about which patients will succeed with NIRS and which require early intubation, a decision whose consequences are severe in either direction.[7] When successful, NIRS reduces the work of breathing and helps prevent complications related to intubation, such as ventilator-associated pneumonia, sedation-induced delirium, and other issues following extubation.[2] However, when NIRS fails, it can delay intubation, increasing the risk of iatrogenic lung injury and progression to more severe illness requiring emergent intervention, all of which are associated with a higher risk of death.[8] The population-level burden is substantial: acute respiratory failure accounts for about 10.4% of ICU admissions globally, with hospital mortality ranging from 35% in mild cases to 46% in more severe cases, a median ventilation duration of 8 days, and a 23-day hospital stay among survivors.[1]

Clinical trials and practice guidelines establish population-level guidance regarding the comparative effectiveness of NIRS modalities. However, these comparisons are typically made against each other or standard oxygen, rather than against IMV.[9] For example, the FLORALI trial compared HFNO against standard oxygen and NIPPV rather than IMV, finding lower 90-day mortality despite a null primary outcome.[10] This gap between population-level trial averages and individual-level heterogeneity motivates the estimation of the individualized treatment effect (ITE). A recent editorial on individualized treatment effect estimation in critical care cautions that machine learning approaches, while capable of discovering complex relationships in observational and trial data, may omit process-level variables that confound the estimated treatment response in individual patients.[11] The meta-learner framework, which includes T-Learners, X-Learners, R-Learners, and doubly robust learners, provides a flexible architecture for outcome modeling and propensity scoring.[12] Causal forests provide consistent estimates of conditional average treatment effects with inference through sample-splitting.[13] The doubly robust learner ensures consistency if either the outcome model or the propensity score is correctly specified,[14] while counterfactual regression networks address treatment selection bias by learning balanced feature representations across treatment groups.[15] However, these approaches were developed for continuous or binary outcomes, not for composite endpoints with competing risks.

Studies applying ITE methods to respiratory support reveal several gaps. A retrospective analysis of 2,354 COVID-19 patients found that initial NIRS was associated with higher mortality risk and earlier discharge, creating a competing-risks pattern that complicates the interpretation of treatment effects.[16] A deep counterfactual inference study comparing HFNO and NIPPV found significant variability in individual-level treatment responses but did not examine combined time-to-event outcomes.[17] None of these studies separated survival from ventilation duration when estimating treatment effects. Digital twin approaches have shown promise in related ICU settings. A deep-learning digital twin for ICU ventilatory support predicts physiological responses to changes in ventilator settings but does not estimate treatment comparisons between support modalities.[18]

Treatment effects are not uniform, and they vary with patients' baseline characteristics. The Predictive Approaches to Treatment Effect Heterogeneity (PATH) statement distinguished risk modeling for predicting baseline disease severity from effect modeling for estimating how patient characteristics modify treatment response.[19] Recent advances in machine learning methods enable flexible estimation of treatment effects using metalearner architectures, causal forests, and generative counterfactuals. These advances have given rise to the twin paradigm, which simulates patient outcomes under different treatments to compute treatment contrasts.[20] Despite this progress, ITE estimation for non-invasive respiratory support remains underdeveloped. Prior ITE approaches for respiratory support have relied on binary or mortality outcomes, disregarding the duration-of-ventilation information that carries substantial clinical weight. Ventilator-free days at 28 days (VFD-28) is a composite endpoint integrating mortality and ventilation duration among survivors by counting days alive and free of mechanical ventilation within 28 days, where a score of zero is assigned to patients who die or remain mechanically ventilated throughout.[21] The 28-day window is standard because most patients have either died or been successfully weaned by day 28, and longer windows increase skew and reduce statistical power.[22] Yet no prior study has addressed the competing-risk structure of VFD-28 in individualized treatment effect estimation, where a score of zero combines death with prolonged mechanical ventilation, two outcomes that are mechanistically distinct and carry fundamentally different implications for treatment benefit. This structure makes the average treatment effect difficult to interpret directly and necessitates methods that explicitly separate survival from ventilation duration.

In this study, we propose a censoring-aware Digital Twin for Individualized Treatment Effects of Non-Invasive Respiratory Support (DINIRS). The DINIRS framework addresses the competing-risk structure of VFD-28 by separating the endpoint into survival probability and conditional ventilation duration. We then integrate doubly-robust pseudo-outcome estimation with counterfactual generation and combine the resulting estimates with tree-based learners to mitigate treatment selection bias inherent in observational ICU data.

## 2.0 Materials and Methods

The DINIRS architecture begins by encoding the first 24 hours of each patient's physiology using a survival-aware transformer that separates mortality and ventilation signals, continues with pretraining the counterfactual generator, and concludes with doubly robust ITE estimation that couples pseudo-outcomes with cross-fitted tree-based learners, as illustrated in **Figure 1**.

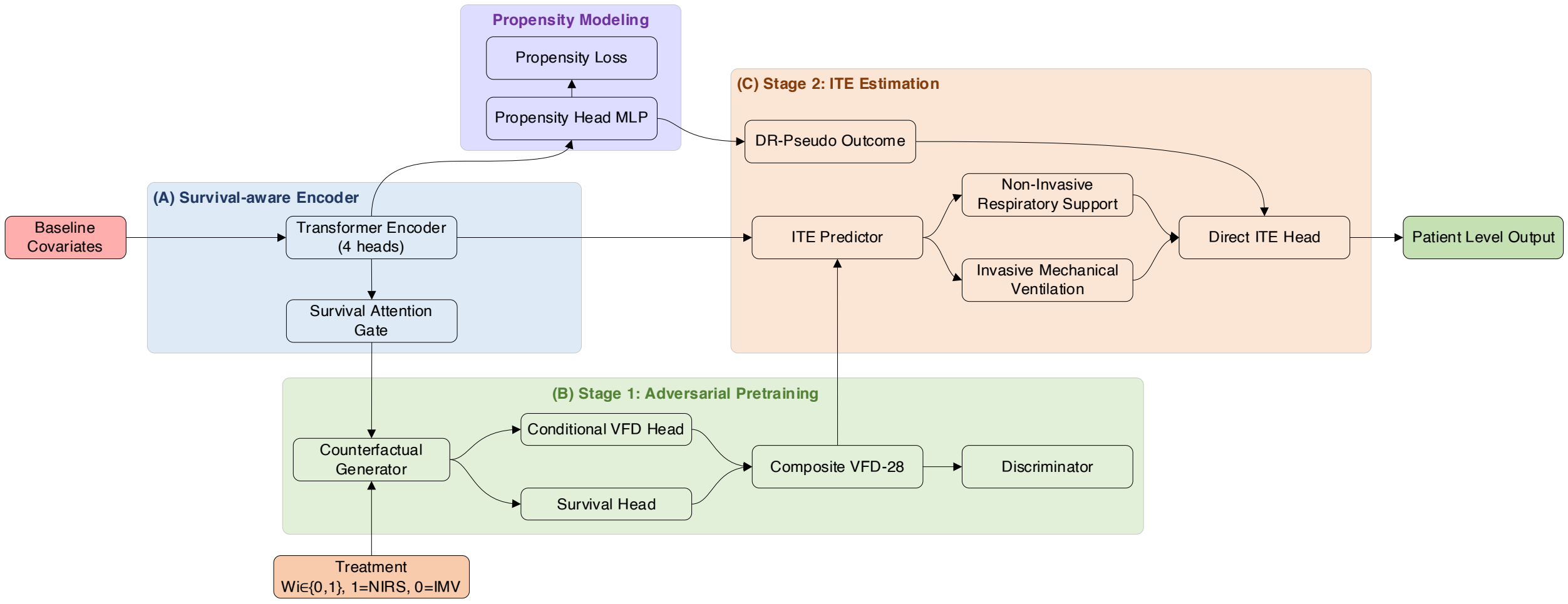

**Figure 1.** Architecture of the DINIRS Framework for Individualized Treatment Effect Estimation. DR: Doubly-Robust; ITE: Individualized Treatment Effect; MLP: Multi-Layer Perceptron; VFD-28: Ventilator-Free Days at 28 Days

## 2.1 Study Design and Data Sources

We used the target trial emulation framework[23] to estimate individual treatment effects of NIRS versus IMV using MIMIC-IV, a publicly accessible critical care database with 94,458 ICU stays.[24] The study focused on adults (≥18 years) admitted with acute respiratory failure who needed respiratory support within 24 hours, with a 28-day follow-up. We excluded patients with "do not resuscitate" orders, those on active mechanical ventilation at admission, or those who died within 24 hours, as these patients had predetermined treatment courses or insufficient observation time to evaluate treatment response.[25] The main outcome was ventilator-free days at 28 days (VFD-28), with secondary outcomes of 28-day mortality and ICU length of stay. We used the full cohort with cross-fitted propensity adjustment and propensity score matching (caliper of 0.1 SD). This created a subset of 1,624 patients for covariate balance assessment and evaluation anchoring. To test generalizability, we created a similarly defined cohort from eICU-CRD, which includes data from 208 US hospitals (2014-2015),[26] resulting in 2,540 patients (630 NIRS and 1,910 IMV).

## 2.2 Covariates

We selected 23 baseline clinical variables from six domains (see **Supplementary Table A1**): demographics, acute illness severity (SOFA, SAPS-II, GCS), vital signs, arterial blood gas parameters, comorbidities, and respiratory indices (P/F ratio, ROX index). We chose these variables based on three principles outlined in the PATH Statement: known influence on respiratory failure outcomes, consideration by clinicians when selecting NIRS or IMV, and availability in both the MIMIC-IV and eICU-CRD datasets for external validation. We used a temporal tensor with 23 physiologic channels, sampled every 30 minutes over the first 24 hours (48-time steps). Missing baseline covariates were imputed using predictive mean matching with five donors.[16,27] To account for imputation uncertainty, we conducted multiple imputation via chained equations with 50 imputations and combined results with Rubin's rules.[28] The FLORALI trial also showed that HFNO efficacy varies across acute respiratory distress syndrome severity levels, and recent findings also highlight nonlinear relationships between $PaCO_2$ and NIPPV failure.[29]

### 2.3 DINIRS Framework Architecture

The DINIRS framework consists of three integrated components. For each patient $i$ with baseline covariates $X_i \in \mathbb{R}^{23}$ and treatment assignment $W_i \in \{0,1\}$ (where $W_i$=1 denotes NIRS and $W_i$=0 denotes IMV), we define potential outcomes $Y_i(0)$ and $Y_i(1)$ as the VFD-28 values that the patient would experience under each treatment. Under the causal assumptions, the expected ITE for patients is:

$$\tau_i = E[Y_i(1) - Y_i(0) \mid X_i] \quad (1)$$

#### 2.3.1 Survival Encoder with Attention Gate

The survival encoder is a 4-layer Transformer encoder[30] with four attention heads, embedded dimension $d_{\text{model}}$=128, feed-forward hidden dimension $d_{\text{ff}}$=256, and dropout probability 0.1. The encoder processes the 48 × 23 temporal covariate tensor $X_i$ and generates a 128-dimensional learned patient representation $z_i \in \mathbb{R}^{128}$. To separate mortality from ventilation duration within this representation, we apply a learned survival attention gate parameterized as a two-layer MLP with sigmoid activation:

$$g_i = \sigma(W_2 \, \text{ReLU}(W_1 z_i + b_1) + b_2) \quad (2)$$

where $W_1, W_2 \in \mathbb{R}^{d_{\text{model}} \times d_{\text{model}}}$ are learned weight matrices and $\sigma(\cdot)$ is the element-wise sigmoid function. This gate produces a vector $g_i \in [0,1]^{d_{\text{model}}}$ that partitions the encoder representation into survival-relevant and ventilation-relevant dimensions: $z_i^{\text{surv}}$=$g_i \odot z_i$ and $z_i^{\text{vfd}}$=(1-$g_i$) $\odot$ $z_i$, where $\odot$ denotes element-wise multiplication. Separate prediction heads then produce survival and conditional ventilation estimates from these separated representations during counterfactual pretraining. The final VFD-28 prediction is:

$$\hat{Y}_i = \hat{S}_i \cdot \hat{V}_i \quad (3)$$

where $\hat{S}_i \in [0, 1]$ is the predicted 28-day survival probability produced from $z_i^{\text{surv}}$ and $\hat{V}_i \in [0,28]$ is the predicted conditional ventilator-free days among survivors produced from $z_i^{\text{vfd}}$. This multiplicative separation directly addresses the fundamental incomparability of the two components: a patient who dies ($\hat{S}$=0) contributes zero ventilator-free days regardless of $\hat{V}$, and a patient who survives but requires prolonged ventilation ($\hat{V} \approx 0$) similarly yields low VFD-28. By learning the gate $g_i$, the model discovers which representation dimensions are informative for mortality prediction versus ventilation duration.

#### 2.3.2 Counterfactual Generator

We adapted the estimation of individualized treatment effects using generative adversarial nets (GANITE) framework[31] for counterfactual outcome generation. Our adaptation required two modifications: (1) our generator operates on the learned encoder representation $z_i$, enabling end-to-end representation learning jointly with counterfactual generation, whereas GANITE operates on raw covariates; (2) we adopted the conditional architecture corresponding to GANITE's ITE estimation block rather than the component-wise counterfactual discriminator. The generator $G_\theta$ takes the patient representation and produces predicted potential outcomes for both arms:

$$[\hat{Y}_i(0), \hat{Y}_i(1)] = G_\theta(z_i) \tag{4}$$

A discriminator network $D_\phi$ learns to distinguish real observed outcomes $Y_i$ from generated counterfactual outcomes, encouraging the generator to produce realistic outcome distributions. Following the conditional adversarial formulation:

$$\mathcal{L}_{\text{adv}} = \mathbb{E}[\log D_\phi(z_i, Y_i)] + \mathbb{E}\left[\log\left(1 - D_\phi\big(z_i, G_\theta(z_i)\big)\right)\right] \tag{5}$$

The generator is trained to minimize -$E[\log D_\phi(z_i, G_\theta(z_i))]$, encouraging generation of outcomes that fool the discriminator.

#### 2.3.3 Doubly-Robust ITE Module

To obtain ITE targets, we computed propensity scores $\hat{\pi}(x_i)$=$P(W_i$=$1|X_i$=$x_i)$ through cross-fitted logistic regression on all 23 baseline covariates. We followed the doubly-robust learner framework[14], where we constructed pseudo-outcomes using the efficient influence function:

$$\hat{\varphi}_i = \hat{\mu}(x_i, 1) - \hat{\mu}(x_i, 0) + \frac{W_i - \hat{\pi}(x_i)}{\hat{\pi}(x_i)\big(1 - \hat{\pi}(x_i)\big)}\big(Y_i - \hat{\mu}(x_i, W_i)\big) \tag{6}$$

In efficient influence function, $\hat{\mu}(x_i, w)$ is the generator's predicted conditional outcome for treatment arm $w$, trained jointly within the adversarial framework. This formulation offers a key benefit: the pseudo-outcome ITE estimates remain consistent if either the propensity model $\hat{\pi}$ or the outcome model $\hat{\mu}$ is correctly specified, offering robustness to misspecification in either part. The ITE predictor network is then trained to backpropagate these pseudo-outcomes onto the learned representations, minimizing the squared loss:

$$\mathcal{L}_{\text{ITE}} = \mathbb{E}[(\tau_\theta(z_i) - \hat{\varphi}_i)^2] \tag{7}$$

We employed the Dragonnet architecture, in which we included a joint propensity head that shares early network layers with the outcome prediction.[32] The total training loss is:

$$\mathcal{L}_{\text{total}} = \mathcal{L}_{\text{ITE}} + \lambda_{\text{adv}}\,\mathcal{L}_{\text{adv}} + \lambda_{\text{MMD}}\,\text{MMD}\big(z^{(1)}, z^{(0)}\big) + \lambda_{\text{prop}}\,\mathcal{L}_{\text{prop}} \tag{8}$$

where $\lambda_{\text{adv}}, \lambda_{\text{MMD}}, \lambda_{\text{prop}}$ are hyperparameter weights. In the reported configuration, the adversarial and maximum mean discrepancy (MMD) weights are set to zero (see **Supplementary Table A2**).

### 2.4 Training Procedure and Evaluation Strategy

The training process consists of two phases: first, pretraining, in which the encoder and generator are trained jointly on observed outcomes, and second, ITE refinement training, in which the doubly-robust predictor is trained using pseudo-outcome targets. All networks are optimized with Adam and use early stopping with a patience of 10 epochs. The MIMIC-trained fold models, scaler, and imputation statistics are applied unchanged to the full eICU-CRD cohort (n=2,540) for external validation without retraining.[33]

To assess the DINIRS framework, we considered four key metrics: discrimination, calibration, clinical impact, and external validity.[34,35] C-for-Benefit evaluates how well the model distinguishes individuals with higher observed benefits by testing whether patient pairs with larger observed benefits also have higher predicted benefits.[36] External validation was performed by applying the trained model to the eICU-CRD dataset without retraining. We then compared DINIRS with three existing ITE estimation methods: T-Learner,[12] Causal Forest,[13] and Causal Survival Forest, which extends the causal forest to right-censored outcomes.[37] Observed practice is reported as a reference policy. Sensitivity to unmeasured confounding is evaluated using E-values and Rosenbaum bounds.[38]

## 3.0 Results

### 3.1 Cohort Characteristics

The analysis cohort comprised 5,336 ICU stays (1,961 NIRS and 3,375 IMV) from the MIMIC-IV cohort, in which both treatment modalities were clinically viable (see **Figure 2**). A separate validation cohort of 2,540 patients (630 NIRS and 1,910 IMV) was extracted from eICU-CRD. The mean age was 62.3 ± 15.4 years, and 58.0% were male. This cohort had moderate-to-high illness severity: the mean Sequential Organ Failure Assessment (SOFA) score was 6.7 ± 3.9, the mean P/F ratio was 192.3 ± 124.9, and 32.8% met sepsis criteria. The cohort mortality within 28 days was 24.0%, and the mean VFD-28 was 19.3 ± 11.5 days (21.2 in the NIRS arm versus 18.2 in the IMV arm). In the propensity-matched subset, all standardized mean differences between treatment groups were below 0.10 (see **Supplementary Tables A1** and **A3**).

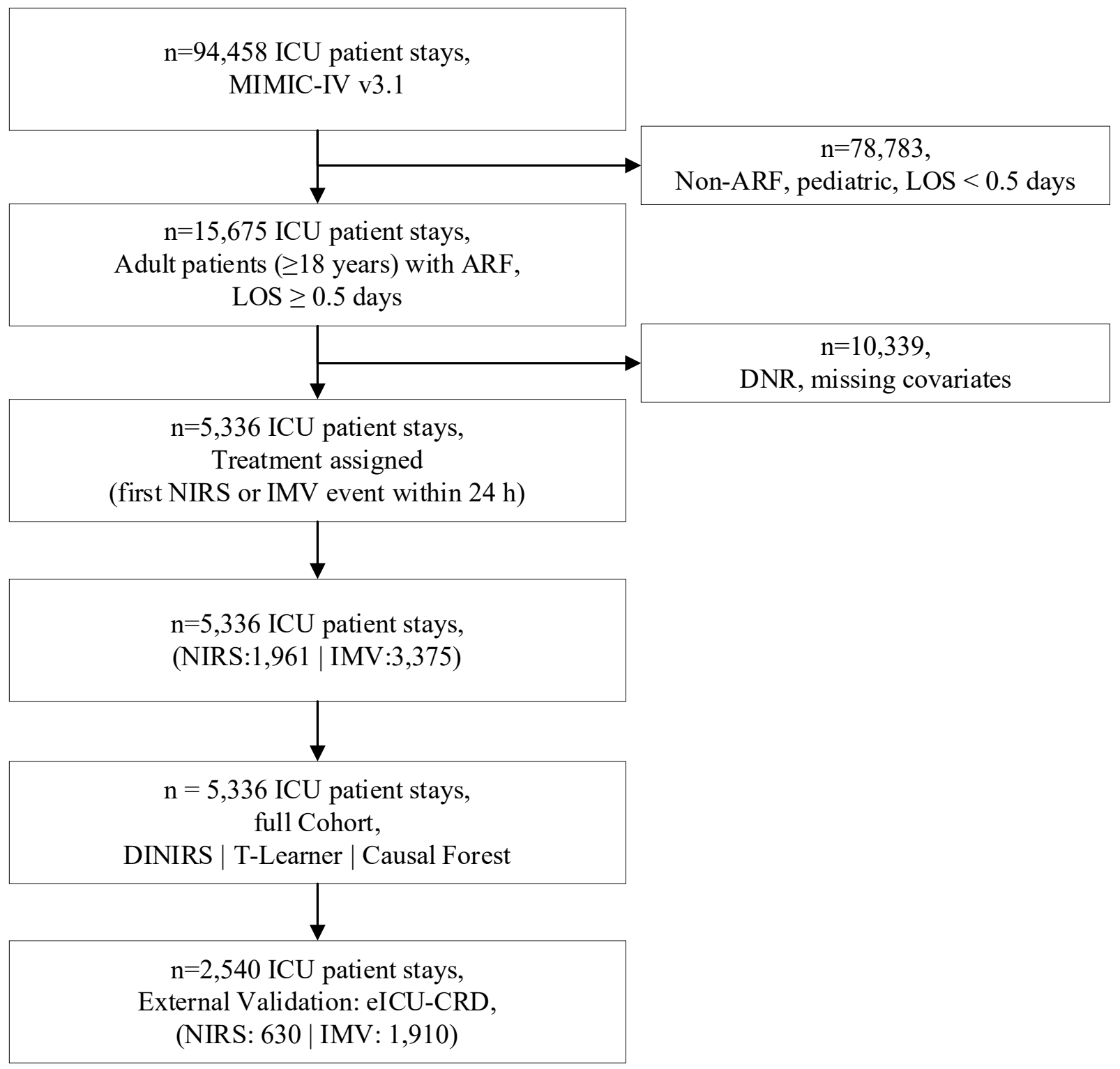

**Figure 2**. Inclusion and exclusion criteria for the analysis cohort.
ARF = acute respiratory failure; LOS = length of stay; DNR = do not resuscitate; NIRS = non-invasive respiratory support; IMV = invasive mechanical ventilation.

### 3.2 Treatment Effect Estimation

The DINIRS model estimated a mean ITE of +2.33 ± 6.05 VFD-28 days for the entire cohort. Of 5,336 patients, 65.3% were classified as NIRS-beneficial (ITE > 0). Individual ITEs ranged from -18.8 to +28.0 days. The survival attention gate analysis revealed that the 28-day survival ITE was 0.015 in probability, whereas the conditional ventilation duration ITE was +1.85 days among survivors. **Figure 3** shows ITEs across clinical subgroups. The largest treatment effects favoring

NIRS were observed in patients with low SOFA scores, whereas patients with high SOFA scores showed ITEs near zero or slightly favoring IMV.

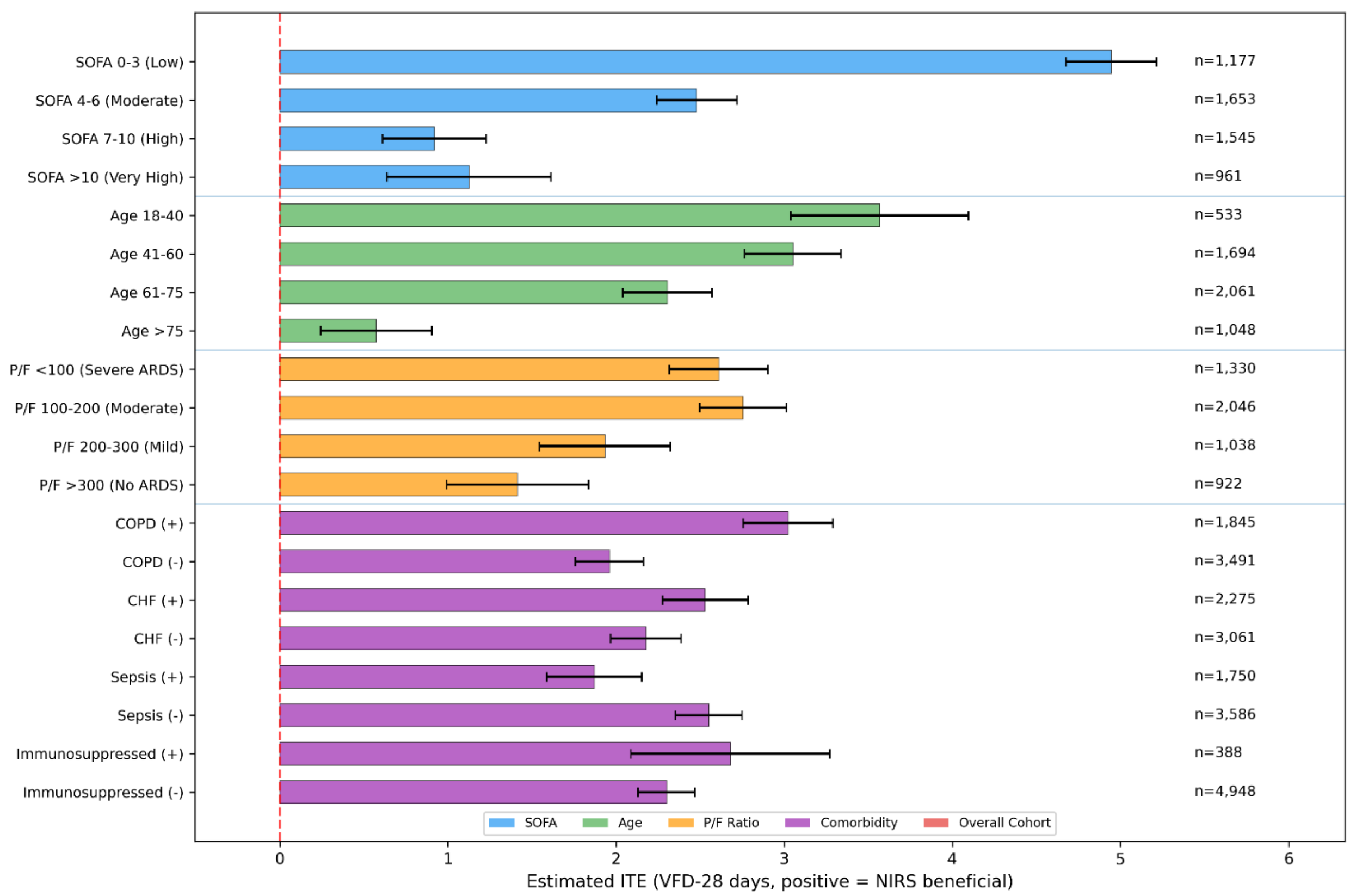

**Figure 3**. Estimated individualized treatment effects (ITEs) across clinical subgroups defined by SOFA score, P/F ratio, age, and comorbidity. Bar length represents the mean ITE, and black horizontal lines indicate 95% confidence intervals. The dashed vertical line at ITE = 0 indicates no treatment effect.

**Table 1** compares DINIRS with T-Learner, Causal Forest, Causal Survival Forest, and observed practice. DINIRS achieved the highest c-for-benefit of 0.621, significantly higher than T-Learner and Causal Forest. DINIRS also achieved the highest doubly robust policy value of 20.87 VFD-28 days, improving by 2.07 days over observed practice (18.80 days). The policy of treating every patient with NIRS could give a VFD of 20.53 days.

**Table 1.** Cross-Method Comparison of ITE Estimation Approaches.

| Method | Mean ITE (SD) | % NIRS-Benefit | Policy Value | C-for-Benefit |
|---|---|---|---|---|
| **DINIRS** | 2.33 (6.05) | 65.3 | 20.87 | 0.621 |
| **T-Learner** | 2.12 (2.82) | 78.3 | 20.67 | 0.600 |
| **Causal Forest** | 3.77 (1.45) | 99.5 | 20.52 | 0.574 |
| **Causal Survival Forest** | 4.02 (1.52) | 99.7 | 20.57 | 0.618 |
| **Observed practice** | - | 36.8 | 18.80 | - |

### 3.3 Clinical Subgroup Analysis

Among patients with SOFA 0-3, 88.4% were predicted to benefit from NIRS, with a mean ITE of +4.95 days (see **Table 2**). The percentage decreased with higher scores, falling to 49.0% for SOFA > 10. A different trend was observed across respiratory indices: 68.8% of patients in the P/F < 100 group were classified as NIRS-beneficial, compared with 54.7% in the P/F > 300 group.

**Table 2.** Subgroup Analysis by SOFA Score Level.

| SOFA Stratum | n (%) | NIRS Benefit (%) | Mean ITE (SD) | Policy Value | C-for-Benefit |
|---|---|---|---|---|---|
| 0-3 | 1,177 (22.1) | 88.4 | 4.95 (4.71) | 24.64 | 0.591 |
| 4-6 | 1,653 (31.0) | 70.3 | 2.48 (4.95) | 22.63 | 0.582 |
| 7-10 | 1,545 (29.0) | 52.6 | 0.92 (6.17) | 20.37 | 0.568 |
| >10 | 961 (18.0) | 49.0 | 1.12 (7.72) | 14.02 | 0.526 |
| Overall | 5,336 (100.0) | 65.3 | 2.33 (6.05) | 20.87 | 0.621 |

**Figure 4** ranks all 5,336 patients by estimated ITE. Patients with SOFA scores of 0-3 (indicated by green bars) are mainly found on the NIRS-beneficial side, whereas those with SOFA > 10 (red bars) appear predominantly toward the IMV-beneficial end (Spearman correlation between SOFA and estimated ITE: -0.29).

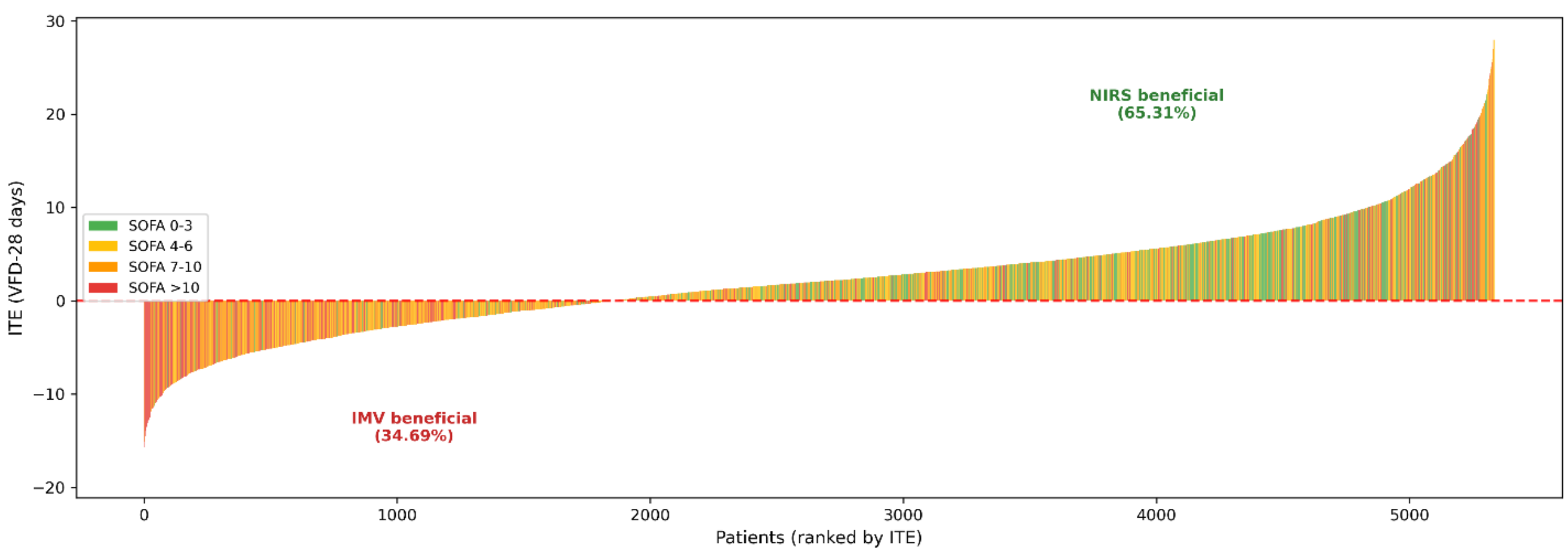

**Figure 4**. Waterfall plot of individualized treatment effects for all 5,336 patients, ranked from most IMV-beneficial to most NIRS-beneficial. Bar colors indicate SOFA levels. The red dashed horizontal line at ITE = 0 separates patients predicted to benefit from NIRS (above) from those predicted to benefit from IMV (below), with around 65.3% of patients predicted to benefit from NIRS.

**Figure 5** illustrates the probability of NIRS benefit across different combinations of SOFA scores and P/F ratios. Cells with low SOFA scores show a 78% to 94% probability of NIRS benefit across the entire range of P/F ratios, whereas cells with SOFA above 6 tend toward an even split.

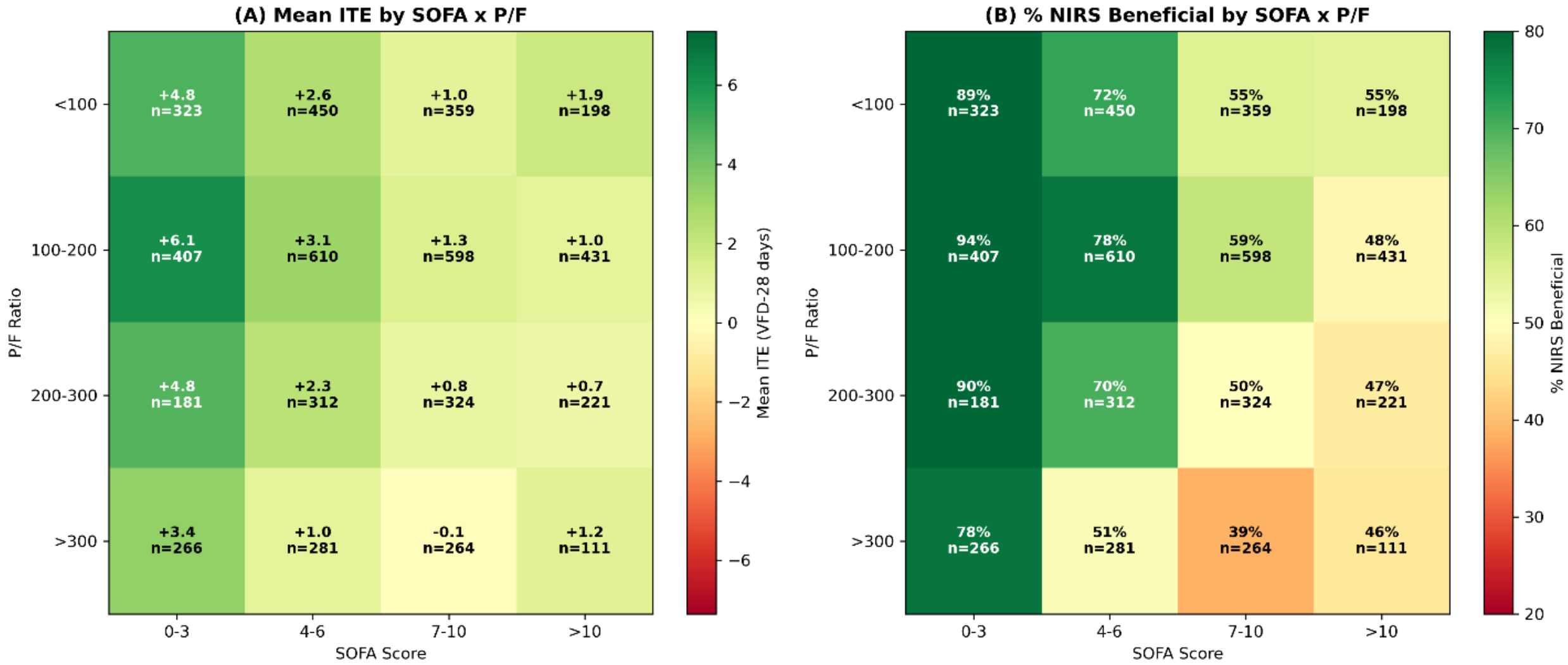

**Figure 5.** Clinical decision support matrix across combinations of SOFA score and P/F ratio. Panel (A): mean predicted ITE (VFD-28 days) per stratum. Panel (B): percentage of patients predicted NIRS-beneficial per stratum.

### 3.4 External Validation and Ablation Study

We applied the pre-trained DINIRS model directly to the eICU-CRD cohort (n=2,540). The model achieved a c-for-benefit of 0.575, with an average treatment effect of +1.94 VFD-28 days. Treatment-effect heterogeneity was replicated: the observed NIRS effect was significantly larger among patients predicted to benefit (+12.48 versus +3.78 days).

We evaluated robustness to unmeasured confounding using E-values and Rosenbaum bounds. The E-value was 2.24, indicating that an unmeasured confounder would need to have a risk ratio of at least 2.24 with both treatment and outcome to account for the estimated average treatment effect. Rosenbaum bounds exceeded 3.0, implying that the direction of treatment-effect heterogeneity is more robust to hidden bias than the point estimates. In the twin ablations, removing the doubly robust pseudo-outcome loss produced the largest performance drop (10.5%), while removing the survival attention gate regularization reduced performance by 4.4%. Adding MMD representation balancing did not improve discrimination (0.566 versus 0.571). Complete ablation results are presented in **Supplementary Table A4**.

### 3.5 Clinical Instances vs Model Recommendations

We presented four patient scenarios to illustrate the range of clinical severity and to compare cases in which the model's recommendation was concordant or discordant with the treatment received (see **Table 3**). A 43-year-old woman with COPD and congestive heart failure but minimal organ dysfunction (Patient 1) was initiated on NIRS, which aligns with the model's recommendation for NIRS benefit. She achieved full ventilator-free survival (VFD-28 = 28.0), all three estimation methods agreed on NIRS benefit, and her covariate profile met guideline criteria for NIPPV in COPD exacerbations and cardiogenic pulmonary edema. Patient 2 had a more severe case with multi-organ dysfunction and depressed consciousness. This patient was intubated, and the model also recommended IMV (ITE -9.6), highlighting that the severity of illness may preclude benefit from NIRS when airway-protective reflexes are impaired. Notably, T-Learner and Causal Forest both predicted a slight NIRS benefit for this patient, illustrating how methods that do not account for the competing-risk structure of VFD-28 can produce conflicting recommendations at high severity.

**Table 3**. Example of Clinical Instances

| | **Patient 1** | **Patient 2** | **Patient 3** | **Patient 4** |
|---|---|---|---|---|
| Scenarios | COPD and CHF, low severity | Multi-organ dysfunction, depressed consciousness | COPD exacerbation | COPD and CHF, elevated lactate |
| Demographics | Female, age 43, BMI of 37.8 | Male, age 64, BMI of 30.0 | Male, age 59, BMI of 30.0 | Male, age 73, BMI of 18.7 |
| SOFA | 1 | 16 | 5 | 9 |
| GCS | 15 | 5 | 14 | 15 |
| SAPS-II | 16 | 101 | 48 | 74 |
| P/F ratio | 113 | 169 | 108 | 214 |
| $SpO_2$ (%) | 92 | 97 | 97 | 97 |
| $PaO_2$ / $PaCO_2$ | 52 / 69 | 98 / 43 | 74 / 49 | 126 / 32 |
| $FiO_2$ (%) | 46 | 58 | 69 | 59 |
| ROX index | 9.8 | 6.3 | 9.0 | 8.5 |
| Lactate | 1.9 | 2.4 | 1.5 | 5.0 |
| Comorbidities | COPD, CHF | CHF, Sepsis | COPD | COPD, CHF |
| Actual treatment | NIRS | IMV | NIRS | IMV |
| Outcome | Survived | Survived | Survived | Died |
| VFD-28 | 28.0 | 20.4 | 28.0 | 0.0 |
| DINIRS ITE | +14.1 | -9.6 | +2.4 | +3.0 |
| T-Learner ITE | +4.3 | +0.4 | +2.7 | +2.6 |
| Causal Forest ITE | +7.6 | +1.2 | +2.6 | +4.5 |
| Recommendation | NIRS | IMV | NIRS | NIRS |
| Concordance | Concordant | Concordant | Concordant | Discordant |

BMI: Body Mass Index; CHF: Congestive Heart Failure; COPD: Chronic Obstructive Pulmonary Disease; $FiO_2$: Fraction of Inspired Oxygen; GCS: Glasgow Coma Scale; IMV: Invasive Mechanical Ventilation; $SpO_2$: Peripheral Oxygen Saturation; VFD-28: Ventilator-Free Days at 28 days.

The remaining two patient scenarios fall in the middle of the severity range. For Patient 3 with a COPD exacerbation and moderate organ dysfunction, NIRS was initiated, consistent with the model's recommendation (ITE +2.4), and full ventilator-free survival was achieved. In contrast, a COPD patient (Patient 4) with congestive heart failure and elevated lactate levels was intubated despite the model predicting a NIRS benefit (ITE +3.0) for this profile. The patient did not survive (VFD-28 = 0). This estimate may have been driven by preserved oxygenation and respiratory indices in this covariate profile, while elevated lactate suggests severity beyond the respiratory system. T-Learner and Causal Forest also preferred NIRS, highlighting that discordant cases require prospective evaluation rather than retrospective adjudication.

## 4.0 Discussion

### 4.1 Clinical Interpretation of Treatment Effect Heterogeneity

This study identified a consistent decrease in the predicted benefit of NIRS across illness severity, as measured by SOFA score. For patients with minimal organ dysfunction (SOFA 0-3), 88.4% were expected to benefit from NIRS, whereas this proportion decreased to 49.0% for those with SOFA scores above 10. Decomposition through the survival attention gate highlights the mechanism behind this pattern: the survival component of the ITE was small (0.015 in probability), indicating that NIRS provided little mortality benefit over IMV. Instead, the overall treatment effect was driven primarily by a shorter ventilation duration among survivors (+1.85 days). This finding indicates that the benefits of NIRS may be driven by reductions in complications associated with intubation and being on a ventilator rather than by preventing death. Tracheal intubation in critically ill patients has complication rates over 40%, including peri-intubation hypotension, cardiac arrest, esophageal intubation, and aspiration.[7] For patients suitable for non-invasive support, avoiding these complications means fewer days on a ventilator, less sedation, and a reduced risk of ventilator-associated pneumonia. Patients with preserved airway reflexes, adequate respiratory drive, and stable hemodynamics can tolerate NIRS even at higher SOFA scores. For these individuals, avoiding intubation-related complications appears to be the main benefit of treatment.

Given the established clinical pattern, a key concern is whether the estimated effects truly reflect treatment effect heterogeneity or are confounded by the treatment selection process. We addressed this concern through four approaches. First, propensity score matching all 23 baseline covariates ensured that matched NIRS patients could be paired with a severity-equivalent IMV patient, with all standardized mean differences below 0.10 (see **Supplementary Table A3**). Second, the doubly-robust pseudo-outcome approach provides consistency if either the outcome model or the propensity score is correctly specified. Third, all reported estimates are out-of-fold and were reproduced exactly by a complete re-run of the pipeline, from data extraction through training. Fourth, the predicted-benefit interaction was reproduced in the independent eICU-CRD cohort. Although these measures do not replace randomization, they collectively reduce the likelihood that the observed pattern is an artifact of a single institution's treatment preferences. The E-value of 2.24 indicates that only a strong unmeasured confounder could fully account for the estimated average effect.

The policy value of 20.87 VFD-28 days represents a 2.07-day improvement over the observed clinical practice of 18.80 days. While this increase is modest overall, it applies to patients near clinical equipoise, where either treatment yields comparable results. Among the 65.3% of patients classified as NIRS-beneficial, the mean predicted benefit was +5.55 VFD-28 days. The C-for-benefit score of 0.621 indicates moderate ability to distinguish the direction of treatment benefit. These results suggest that DINIRS identifies a clinically meaningful subgroup that could benefit from NIRS, even though population-level averages mask this heterogeneity.

### 4.2 Prior Evidence and Broader Applicability

The pattern of decreasing NIRS benefit from increasing organ dysfunction observed in this study aligns with prior trial evidence and guidelines. Clinical guidelines recommend NIPPV for COPD exacerbations and cardiogenic pulmonary edema, consistent with the model's predictions for these conditions.[3,39] The 2026 American Thoracic Society guideline recommends noninvasive support without restricting eligibility based on hypoxemia severity, consistent with our finding that predicted benefit persisted across the P/F ratio (68.8% below 100 versus 54.7% above 300).[9] The guideline also identifies populations that benefit most as a research priority, though our comparison is with invasive ventilation rather than with standard oxygen therapy. The FLORALI trial demonstrated that HFNO reduced 90-day mortality in acute hypoxemic respiratory failure which is consistent with the predicted benefit at low P/F ratios in our cohort.[10] However, population-level trial results can obscure individual-level heterogeneity. The BOUGIE trial illustrates this, as its primary analysis showed no overall benefit of bougie-assisted intubation,[40] yet post hoc causal forest analyses revealed that patient characteristics, particularly difficult airway features, were the strongest predictors of individualized benefit. This aligns with the broader concern that measured confounders in observational and trial data may not fully capture the mechanisms driving treatment response in individual patients.[11] DINIRS addresses this limitation by shifting from population averages to patient-specific estimates and explicitly modeling the outcome's competing-risk structure.

The survival attention gate addresses a problem not handled by current ITE estimators: competing risks in composite outcomes such as VFD-28, where death and prolonged ventilation both yield a score of zero. The ablation study showed that the doubly-robust pseudo-outcome component made the largest contribution to performance, highlighting the value of combining outcome modeling with inverse propensity weighting, especially when both models are prone to misspecification.[14] These components are not specific to respiratory support; any composite endpoint where a zero score can arise from clinically distinct pathways could benefit from a similar framework.

### 4.3 Limitations and Future Directions

This study has several limitations. First, NIRS in observational data is not a standardized intervention but a clinician-directed process involving modality selection, titration, monitoring, and failure recognition.[4] The 23 covariates capture only baseline patient characteristics, not process variables such as titration trajectory or clinician experience. Second, NIRS is treated as a single intervention containing HFNO, CPAP, and BPAP, each with distinct physiological mechanisms. A patient benefiting from BPAP for COPD exacerbation and one benefiting from HFNO for pneumonia are both classified as NIRS-beneficial, despite different underlying effects. Third, although the eICU-CRD validation suggests the model can generalize across 208 hospitals, the decline in c-for-benefit from 0.621 to 0.575 on external data indicates site-level variation that the model does not fully capture. Fourth, the model estimates ITEs only at the initial support decision point and does not account for changes over time in NIRS trial duration, escalation, or the timing of the transition from NIRS to IMV.[8] Fifth, because DINIRS is trained on historical treatment data, its predictions might reinforce existing clinician biases rather than identify the truly best treatment, leading to high precision but not necessarily accuracy.[11] Factors not measured, such as operator experience, team dynamics, and the patient's real-time progress, fall outside the scope of baseline covariates. As with all machine learning approaches, unusual associations or deviations from performance - in addition to selection bias in the dataset - can be amplified by the algorithms. Additionally, prospective validation of machine learning predictions in a clinical trial is impractical without advances in clinical trial capabilities to stratify randomization based on algorithm predictions. As such, all algorithms aimed at predicting ITE in observational or clinical trial datasets should be viewed as developmental and hypothesis generating.

Future research should focus on estimating modality-specific ITEs for NIRS separately, which would better align recommendations with clinicians' pathophysiology-based choices. Sequential counterfactual frameworks such as G-Net,[41] which employ g-computation for time-dependent treatments, provide a way to model these evolving regimes and to address when and how to adjust respiratory support during an ICU stay. Validating the model prospectively against standard clinical decisions is essential to verify whether the predicted pattern leads to better patient outcomes.

## 5.0 Conclusion

This study introduced the DINIRS framework to estimate individualized treatment effects of non-invasive respiratory support. The model identified a trend in which predicted NIRS benefit was concentrated among patients with lower organ dysfunction and was positive across the range of hypoxemia severity. This pattern aligns with prior trial evidence and clinical guidelines and was confirmed in an independent multi-site cohort from eICU-CRD. The survival attention gate showed that the benefit came mainly from reduced ventilation duration among survivors rather than from a reduction in mortality, pointing to the avoidance of intubation-associated complications as the primary mechanism. Prospective validation with modality-specific ITE estimation is needed before this framework can inform clinical practice.

## 6.0 Data Availability

The DINIRS framework files needed to reproduce the results are available at

https://github.com/vsubbian/DINIRS

## Supplementary Materials

**Table A1** shows the baseline characteristics of the full MIMIC-IV analysis cohort (n=5,336) by treatment arm.

**Table A1.** Baseline Characteristics of MIMIC-IV Cohort (n=5,336).

| Characteristic | Overall (n=5,336) | NIRS (n=1,961) | IMV (n=3,375) | Std. Diff. |
|---|---|---|---|---|
| **Demographics** | | | | |
| Age, years | 62.3 ± 15.4 | 64.2 ± 15.1 | 61.1 ± 15.5 | 0.199 |
| Male, % | 58.0 | 54.3 | 60.1 | 0.118 |
| sBMI, kg/m2 | 30.2 ± 8.7 | 30.6 ± 9.0 | 30.0 ± 8.5 | 0.072 |
| **Severity Scores** | | | | |
| SOFA score | 6.7 ± 3.9 | 4.8 ± 3.1 | 7.8 ± 3.9 | 0.862 |
| GCS score | 13.4 ± 3.1 | 13.8 ± 2.1 | 13.1 ± 3.5 | 0.244 |
| SAPS-II | 42.3 ± 15.4 | 36.3 ± 12.8 | 45.7 ± 15.8 | 0.654 |
| **Vital Signs (24h mean)** | | | | |
| Heart rate, bpm | 88.4 ± 16.7 | 89.2 ± 16.3 | 87.9 ± 17.0 | 0.077 |
| RR, breaths/min | 21.5 ± 4.3 | 22.3 ± 4.5 | 21.0 ± 4.2 | 0.280 |
| SpO2, % | 96.2 ± 2.5 | 94.8 ± 2.1 | 97.0 ± 2.3 | 1.003 |
| MAP, mmHg | 79.0 ± 10.3 | 81.7 ± 11.0 | 77.4 ± 9.6 | 0.420 |
| Temperature, C | 36.9 ± 0.5 | 36.9 ± 0.4 | 36.9 ± 0.6 | 0.128 |
| **Blood Gas (24h mean)** | | | | |
| PaO2, mmHg | 98.5 ± 56.9 | 73.7 ± 42.0 | 113.0 ± 59.5 | 0.762 |
| PaCO2, mmHg | 46.4 ± 12.4 | 49.4 ± 15.1 | 44.6 ± 10.0 | 0.375 |
| pH | 7.35 ± 0.07 | 7.37 ± 0.07 | 7.34 ± 0.07 | 0.428 |
| FiO2 | 56.2 ± 18.0 | 62.9 ± 21.4 | 52.3 ± 14.2 | 0.580 |
| Lactate, mmol/L | 2.4 ± 2.1 | 1.8 ± 1.0 | 2.7 ± 2.4 | 0.528 |
| HCO3, mEq/L | 23.2 ± 5.6 | 25.2 ± 5.9 | 22.0 ± 5.1 | 0.592 |
| **Respiratory Indices** | | | | |
| P/F ratio, mmHg | 192.3 ± 124.9 | 134.1 ± 96.1 | 226.1 ± 127.2 | 0.817 |
| ROX index | 9.4 ± 4.1 | 8.4 ± 4.5 | 10.0 ± 3.7 | 0.391 |
| **Comorbidities, %** | | | | |
| COPD, % | 34.6 | 44.4 | 28.9 | 0.325 |
| CHF, % | 42.6 | 49.7 | 38.5 | 0.225 |
| Immunosuppressed, % | 7.3 | 9.8 | 5.8 | 0.152 |
| Sepsis-3, % | 32.8 | 30.9 | 33.9 | 0.066 |

| Outcomes | | | | |
|---|---|---|---|---|
| VFD-28, days | 19.32 ± 11.52 | 21.21 ± 11.38 | 18.23 ± 11.46 | 0.261 |
| 28-day mortality, % | 24.0 | 21.2 | 25.6 | 0.104 |

Values are mean ± SD or percentage unless otherwise noted. Std. Diff.: Absolute standardized mean difference before propensity score matching.

**Table A2** shows the hyperparameters used for training DINIRS Architecture.

**Table A2:** Hyperparameters for DINIRS Architecture and Training.

| Component | Parameter | Value |
|---|---|---|
| Encoder | Architecture | Transformer, 4 layers |
| | Embedding dimension ($d_{\text{model}}$) | 128 |
| | Attention heads | 4 |
| | Feed-forward dimension ($d_{\text{ff}}$) | 256 |
| | Dropout probability | 0.10 |
| Generator | Architecture | 3-layer MLP |
| | Hidden dimensions | [256, 128] |
| | Output dimension | 2 (two treatment arms) |
| | Activation | ReLU |
| Discriminator | Architecture | 2-layer MLP |
| | Hidden dimension | 64 |
| | Output dimension | 1 (binary) |
| Optimization | Optimizer | Adam |
| | Learning rate | $1 \times 10^{-4}$ |
| | Batch size | 128 |
| | Early stopping patience | 10 epochs |
| | Stage 1 epochs | 100 |
| | Stage 2 epochs | 50 |
| Loss Weights | $\lambda_{\text{adv}}$ | 0 (ablation-guided) |
| | $\lambda_{\text{MMD}}$ | 0 (ablation-guided) |
| | $\lambda_{\text{prop}}$ | 1.0 |
| Other | PSM caliper | 0.1 SD |
| | Train / test split | 5-fold cross-fitting (out-of-fold) |
| | Propensity estimation | Logistic regression |

**Table A3** shows the standardized mean differences for 23 covariates before and after propensity score matching. All post-match SMDs weres under 0.10, indicating good balance.

**Table A3.** Standardized Mean Differences Before and After Propensity Score Matching.

| Covariate | Pre-Match SMD | Post-Match SMD |
|---|---|---|
| Age | 0.199 | 0.027 |
| Male | 0.118 | 0.017 |
| BMI | 0.072 | 0.013 |
| SOFA | 0.862 | 0.010 |
| GCS | 0.244 | 0.011 |
| SAPS-II | 0.654 | 0.032 |
| Heart rate | 0.077 | 0.000 |
| RR | 0.280 | 0.025 |
| SpO2 | 1.003 | 0.003 |
| MAP | 0.420 | 0.007 |
| Temperature | 0.128 | 0.003 |
| PaO2 | 0.762 | 0.017 |
| PaCO2 | 0.375 | 0.023 |
| pH | 0.428 | 0.022 |
| FiO2 | 0.580 | 0.010 |
| Lactate | 0.528 | 0.007 |
| Bicarbonate (HCO3) | 0.592 | 0.011 |
| P/F ratio | 0.817 | 0.005 |
| ROX index | 0.391 | 0.005 |
| COPD | 0.325 | 0.005 |
| CHF | 0.225 | 0.032 |
| Immunosuppressed | 0.152 | 0.028 |
| Sepsis-3 | 0.066 | 0.008 |

**Table A4** presents the results of ablation studies quantifying the contribution of each architectural component. Ablations are evaluated on the twin components alone (pooled out-of-fold), and the reported configuration achieves a c-for-benefit of 0.571. Table 1 reports a c-for-benefit of 0.621 for the full stacked estimator.

**Table A4.** Ablation Study: Contribution of DINIRS Components (c-for-benefit, out-of-fold, n=5,336).

| **Model Variant** | C-for-Benefit (95% CI) | ITE SD |
|---|---|---|
| Twin alone | 0.571 (0.557 to 0.589) | 3.82 |
| Without survival attention gate | 0.546 (0.535 to 0.566) | 2.93 |
| Without DR pseudo-outcome loss | 0.511 (0.496 to 0.526) | 0.67 |
| With MMD balancing added | 0.566 (0.551 to 0.582) | 3.23 |
| With adversarial objective added | 0.588 (0.569 to 0.600) | 5.34 |

Compared with the Twin model, neither adding MMD balancing (-0.005, $p = 0.34$) nor adding an adversarial objective (+0.017, $p = 0.17$) produced a statistically significant difference. Therefore, both weights are set to zero in the final configuration. In contrast, removing the survival attention gate (-0.025, $p < 0.001$) and the doubly robust pseudo-outcome loss (-0.060, $p < 0.001$) significantly worsened performance.